\documentclass[cits]{PoS}
\usepackage{bbm}
\usepackage{epsfig}
\title{Strangeness and charm content of the nucleon}

\ShortTitle{Strangeness and charm content of the nucleon}

\author{\speaker{Gunnar Bali}, Sara Collins and Andreas Sch\"afer\\
        Institut f\"ur Theoretische Physik, Universit\"at Regensburg,\\
        93040 Regensburg, Germany\\
        E-mail: \email{gunnar.bali@physik.uni-regensburg.de},
\email{sara.collins@physik.uni-regensburg.de},
\email{andreas.schaefer@physik.uni-regensburg.de}\\\vspace*{.25cm}}

\author{
\rm\centering (QCDSF Collaboration)}

\abstract{We present results on the scalar strangeness
and charm contents of the nucleon and of the
disconnected contributions to the nucleon spin.
These are
obtained on $n_F=2$ non-perturbatively improved
Sheikholeslami-Wilson configurations at a pseudoscalar mass
of 290\,\,MeV. We quote
$f_{T_s}=m_s\langle N|\bar{s}s|N\rangle/m_{\rm N}=0.070(22)$
as our preliminary value for
the strange quark fraction of the nucleon mass
and an $\overline{MS}$ scheme value $\Delta s=-0.015(10)$,
with as yet unknown systematics,
for the
strangeness contribution to the spin.
}
\FullConference{The XXVI International Symposium on Lattice Field Theory - LAT2009\\
                 July 26-31 2009\\
                 Peking University, Beijing, China}
\begin{document}

\section{Introduction}

We calculate disconnected scalar matrix elements
$\langle N|\bar{q}q|N\rangle$ as well as contributions to the
spin  of the nucleon $\Delta q$. 
Of particular interest are currents
containing the strange quark $q=s$,
where only quark-line disconnected diagrams contribute but we
also consider heavier and lighter flavours.

The scalar matrix element determines the coupling of the
Standard Model Higgs boson to the quarks within the proton.
This then might couple to heavy particles that
could be discovered in future LHC experiments, some of
which are dark matter candidates~\cite{Ellis:2009ai}.
The fraction of the proton mass $m_{\rm N}$
carried by
the quark flavour $q$, $f_{T_q}=m_q\langle N|\bar{q}{q}|N\rangle/m_{\rm N}$,
is particularly important
since the combination $m_{\rm N}\sum_qf_{T_q}$ will
appear quadratically in this cross section. It is evident that
$f_{T_q}\rightarrow 0$
for $m_q\rightarrow 0$
and $f_{T_q}\propto \langle N|GG|N\rangle/m_{\rm N}=f_{T_G}$ for
$m_q\rightarrow \infty$. We calculate the
scalar matrix element for quark masses up to the charm quark
to confirm this limiting behaviour. 
It is also not immediately obvious whether the charm quark
should be considered as a sea quark or not.
We remark
that in general there will be mixing between
the dimension four operators $GG$ and $m_q\bar{q}q$.
This deserves future study.

Disconnected contributions to the nucleon structure
are also needed with respect to
precision measurements of Standard Model parameters in $pp$ collisions
at the LHC where for instance the
resolution of a (hypothetical) mass difference between
the $W^+$ and $W^-$ bosons is limited by the theoretical knowledge
of the asymmetries between up and down as well as between strange and
charm sea quark contents of  the proton~\cite{Aad:2009wy}.

The spin of the nucleon can be factorized into
a quark spin contribution
$\Delta\Sigma$, a quark angular momentum contribution $L_q$
and a gluonic contribution (spin and angular momentum) $\Delta G$:
\begin{equation}
\frac12=\frac12 \Delta\Sigma+L_q+\Delta G\,.
\end{equation}
In the na\"{\i}ve non-relativistic
$\mathrm{SU}(6)$ quark model, $\Delta \Sigma=1$, with
vanishing angular momentum and gluon contributions. In this
case there will also be no strangeness
contribution $\Delta s$ in the factorization,
\begin{equation}
\Delta\Sigma=\Delta d+\Delta u +\Delta s+\cdots\,,
\end{equation}
where in our notation $\Delta q$ contains both, the spin of
the quarks $q$ and of the antiquarks $\bar{q}$.
Experimentally, $\Delta s$ is
obtained by integrating the strangeness contribution to
the spin structure function $g_1$ over the momentum fraction $x$.
The integral over the range in which data exists ($x\gtrsim 0.004$)
usually agrees with zero. For instance a recent Hermes measurement
in the region
$x\geq 0.02$ yields~\cite{Airapetian:2008qf} $\Delta s=0.037(19)(27)$.
This means that non-zero results
rely on extrapolations into the experimentally un-probed region of 
very small $x$ and are
model dependent~\cite{Zhu:2002tn,deFlorian:2009vb}.
The standard Hermes analysis~\cite{Airapetian:2007mh}
yields $\Delta s=-0.085(13)(8)(9)$
in the $\overline{MS}$ scheme.

We reported first results and developed
the necessary methods in
refs.~\cite{lat07,Bali:2008sx,Bali:2009hu}.
These algorithmic studies were performed on rooted staggered
$n_F\stackrel{?}{=}2+1$ configurations and indicated a
value, $\Delta s>-0.02$.
However, there are unresolved theoretical issues with this
fermion approach~\cite{Creutz:2007yg},
so that for the physics study, of which
we present preliminary results here, we employ improved Wilson
fermions that have a meaningful continuum limit.

\section{Simulation details}
Simulations were performed on
$24^3\times 48$ and $32^3\times 64$ QCDSF configurations of
$n_F=2$ non-perturbatively improved
clover Wilson
fermions with Wilson gauge action at $\beta=5.29$ and
$\kappa_{\rm sea}=0.13632$. These values correspond to a pseudoscalar
mass $m_{\rm PS}\approx 290\,\,$MeV and a lattice spacing
$a^{-1}\approx 2.59\,\,$GeV~\cite{G\"ockeler:2008we}. The scale
was set from the value
$r_0^{-1}\approx 422\,\,$MeV, obtained by chirally extrapolating
the combination $m_{\rm N}r_0$ to the physical point.
The spatial lattice extents correspond to $La\approx 1.83\,$fm and
$La\approx 2.43\,$fm, with
a larger $40^3\times 64$ volume ($La\approx 3.04\,\mbox{fm}\approx 4.5\,
m_{\rm PS}$) still being analyzed.
We use a
modified version of the Chroma code~\cite{chroma}.

We vary the quark mass parameter $\kappa_{\rm loop}$
of the current insertion as well
as that of the nucleon valence quarks $\kappa_{\rm val}$.
In particular we combine the values $\kappa=\kappa_{\rm sea}=0.13632$,
$\kappa=0.13609$ and $\kappa=\kappa_{\rm strange}=0.13550$,
corresponding to pseudoscalar masses
$m_{\rm PS}\approx 290,\,440$ and 690\,\,MeV, respectively.
Additional heavier masses are used for the loop.

$\Delta q$ and 
$\langle N|\bar{q}q|N\rangle$
are extracted from the ratios of
three-point functions to two-point functions~(at zero momentum):
\begin{equation}
\label{eq:rati}
R^{\rm dis}(t,t_{\rm f}) = 
-
\frac{\mathrm{Re}\,\left\langle\Gamma_{\rm 2pt}^{\alpha\beta}C^{\beta\alpha}_{\rm 2pt}(t_{\rm f}) \sum_{\mathbf{x}}\mathrm{Tr}\,(M^{-1}(\mathbf{x},t;\mathbf{x},t)\Gamma_{\rm loop})\right\rangle_{\!\!c}}{\left\langle \Gamma_{\rm unpol}^{\alpha\beta} C^{\beta\alpha}_{\rm 2pt}(t_{\rm f})\right\rangle_{\!\!c}}\,.
\end{equation}
For the scalar matrix element we use,
$\Gamma_{\rm 2pt}=\Gamma_{\rm unpol}:=(1+\gamma_4)/2$ and
$\Gamma_{\rm loop}=\mathbbm{1}$. For
$\Delta q$ we calculate the difference between
two polarizations:
$\Gamma_{\rm 2pt}= \gamma_j\gamma_5\Gamma_{\rm unpol}$ and
$\Gamma_{\rm loop}=\gamma_j\gamma_5$, where we average over
all three possible $j$-orientations. The spin projection
operators along the $j$-axis
read, $P_{\uparrow\downarrow}=\frac12(\mathbbm{1}\pm i\gamma_j\gamma_5)$,
so that in this case,
$\Gamma_{\rm 2pt}= -i(P_{\uparrow}-P_{\downarrow})\Gamma_{\rm unpol}$,
where we have traded a factor $i$ against taking the real part,
rather than the imaginary part, of the nominator
in eq.~(\ref{eq:rati}).
The variance of the above expression is reduced by
explicitly using the fact that $\mathrm{Im}\,\mathrm{Tr}\,(M^{-1}{\mathbbm 1})
=\mathrm{Im}\,\mathrm{Tr}\,(M^{-1}\gamma_j\gamma_5)=0$.

$C^{\alpha\beta}(t_{\rm f})$ denotes the two-point function 
of the zero momentum projected proton with sink and source
spinor indices $\alpha$ and $\beta$ and positions $t_{\rm f}$
and $t_{\rm i}=0$.
In the limit of large times, $t_{\rm f}\gg t\gg 0$, in the axial case,
\begin{equation}
\label{eq:tlimit}
R^{\rm dis}(t,t_{\rm f})+R^{\rm con}(t,t_{\rm f}) \longrightarrow
\Delta q\,,
\end{equation}
where we have not computed the connected contribution
$R^{\rm con}$. This vanishes for strangeness and charm contents.
In the scalar case
the vacuum contribution 
$\langle 0|\bar{q}q|0\rangle=
-\sum_{\mathbf{x}}
\mathrm{Re}\left\langle\mathrm{Tr}\,(M^{-1}(\mathbf{x},t;\mathbf{x},t)
\right\rangle_{\!c}$ needs to be subtracted from eq.~(\ref{eq:rati}).
We employ sink and source smeared two point functions such that
this asymptotic limit is effectively reached for $t=4a\approx 0.3\,$fm and
$t_{\rm f}=8a\approx 0.61\,$fm. We vary $t_{\rm f}$ to check this
assumption and
compute the final results from the $t_{\rm f}\geq 8a$ data.

The clover Wilson operator $M$ can be written as
$2\kappa M={\mathbbm 1}-\kappa D$. We estimate
$\mathrm{Tr}\,[M^{-1}\Gamma]$
stochastically at the cost of less than 100 preconditioned CG
solves on each configuration. The noise is reduced by
calculating the first two
terms of the hopping parameter expansion explicitly, which
corresponds to multiplying the estimates by $(\kappa D)^2$.
For details, see~\cite{Bali:2009hu} and references therein.

\DOUBLEFIGURE{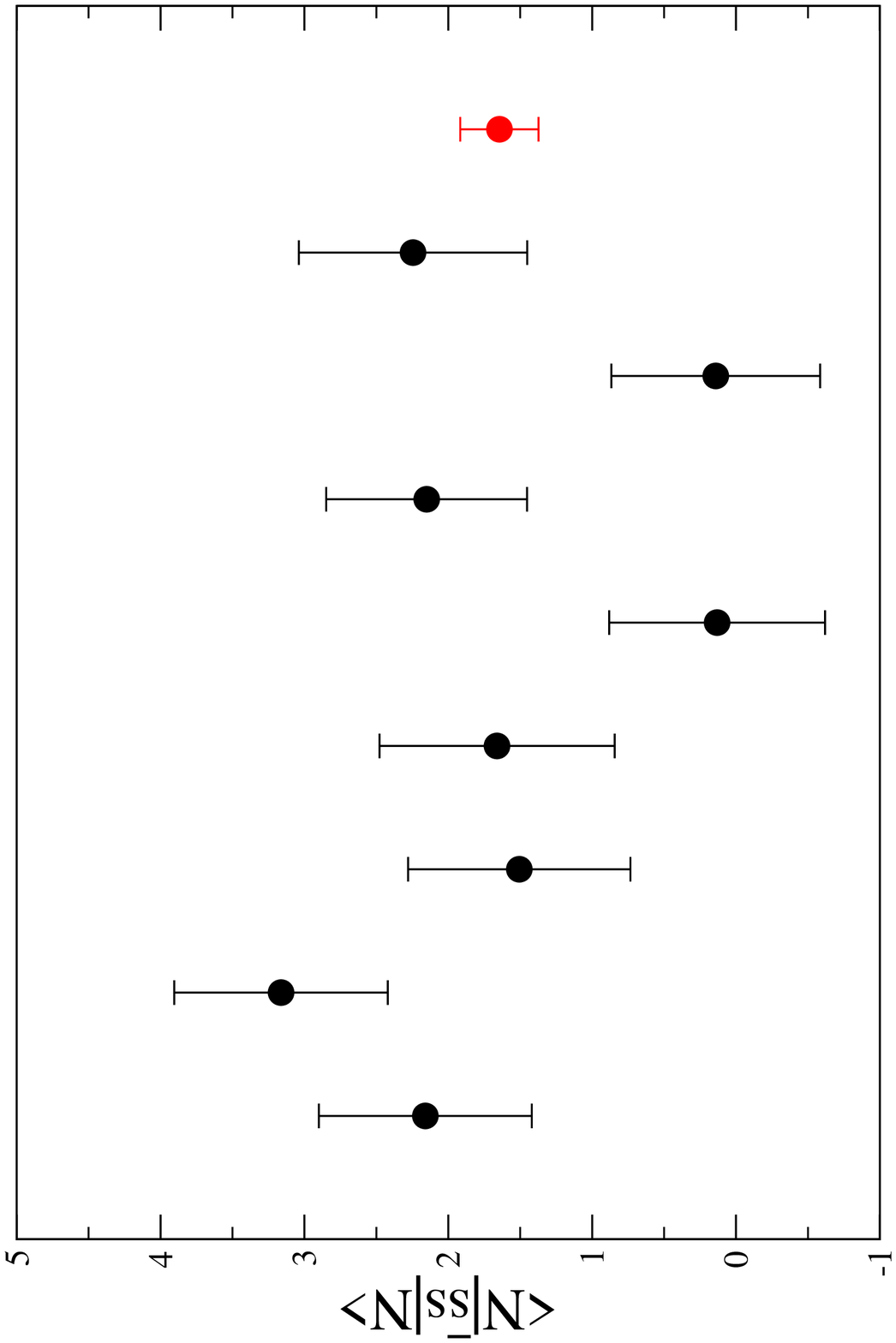,angle=270,width=.46\textwidth,clip=}{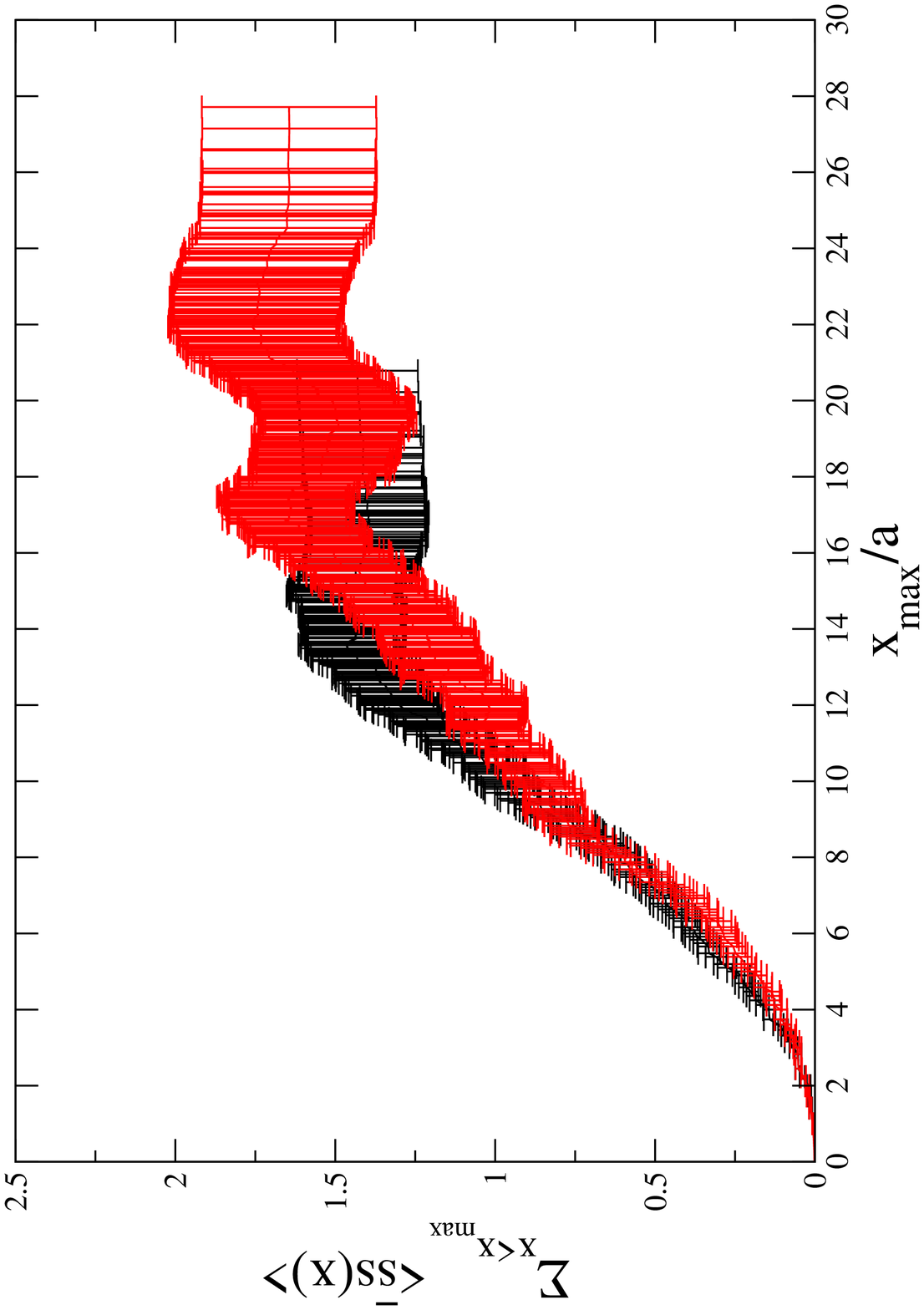,angle=270,width=.48\textwidth,clip=}{The scalar matrix element at
$\kappa_{\rm loop}=\kappa_{\rm val}=\kappa_{\rm strange}$ for 8
different sources and the averaged result.
\label{fig2}}{The partial sum eq.~(\protect\ref{eq:partial})
for the scalar matrix element at
$\kappa_{\rm loop}=\kappa_{\rm val}=\kappa_{\rm strange}$
for $L=32$ (red symbols) and $L=24$ (black symbols).
\label{fig3}}

For the calculation of $\Delta q$
we employ the truncated solver method~\cite{lat07,Bali:2009hu} where for
the strange mass on the $32^3\times 64$ lattice,
$N_1=730$, $N_2=50$, $n_{\rm t}=40$. 
For the truncated solves we use the smoothly converging
even-odd preconditioned CG algorithm
while for the runs to full convergence
we employ the faster BiCGstab2 algorithm.
For the scalar matrix element we only use these 50 solutions since this
is dominated by the gauge noise, not by the stochastic noise.
We deviate from ref.~\cite{Bali:2009hu} by seeding the stochastic sources
on 8 timeslices rather than on one. This does not cause
any computational overhead. We then compute the
standard point-to-all propagators that are needed for the
nucleon two point functions  for 4 time-separated source points
on each configuration
(requiring 48 solves rather than 12). Exploiting forward and
backward 
propagation in time
(replacing
$\Gamma_{\rm 2pt}=(1+\gamma_4)/2$ by $(1-\gamma_4)/2$), this then gives us 8 
two point (and three point) functions
per configuration. Indeed, this averaging reduces the errors of $\Delta s$
and $\langle N|\bar{s}{s}|N\rangle$ by factors $\sim 1/\sqrt{8}$
which is the maximal possible gain. We display this for the
latter example in figure~\ref{fig2}.

In the ongoing analysis of the $40^3\times 64$ volumes we
will compute the lowest eigenmodes of the Hermitian Dirac
operator $\gamma_5M$,
to further reduce the stochastic noise of
the disconnected
loops~\cite{Bali:2009hu}, to precondition the solver and for low mode
averaging of the nucleon two point functions~\cite{Takeda:2009ga}.

In the present study of the $L=24$ and $L=32$ lattices,
where no zero momentum projection is performed at the source,
we find it worthwhile to rearrange
the nominator within eq.~(\ref{eq:rati}):
\begin{equation}
\label{eq:partial}
\sum_{|{\mathbf x}|\leq x_{\rm max}}\left\langle
\sum_{\mathbf y}C_{\rm 2pt}({\mathbf y},t_{\rm f};{\mathbf 0},0)
\,\mathrm{Tr}\,\left[M^{-1}({\mathbf x},t;{\mathbf x},t)\Gamma\right]
\right\rangle_{\!\!\!c}=\sum_{|{\mathbf x}|\leq x_{\rm max}}\!\!\!f({\mathbf x})\sim
x_{\max}^3e^{-mx_{\max}}+\cdots\,.
\end{equation}
At large $x$ the $f(x)\sim e^{-mx}$ values
will eventually not contribute to the signal
anymore but just increase the statistical noise.
We display the partial sums for the 1.8\,fm ($L=24$) and
2.4\,fm ($L=32$) lattices for the scalar matrix element at
$\kappa_{\rm val}=\kappa_{\rm loop}=\kappa_{\rm strange}$
in figure~\ref{fig3}. Indeed, at small $x_{\max}$ we see
the expected $x_{\max}^3$ volume scaling. This flattens somewhat
around $x_{\max}\approx 8a$ but only saturates when the
boundaries of the box are hit ($x_{\max}=12a$ and
$x_{\max}=16a$, respectively). Beyond these distances only the
lattice ``corners'' are summed up. This non-saturation
means that partial sums
can only become a permissible method of reducing the noise at
much larger spatial volumes. It also indicates that finite size
effects might still be substantial for the 2.4\,fm data.
The partial sums appear to saturate even more
slowly for the case of $\Delta s$.

\section{Results}

\DOUBLEFIGURE{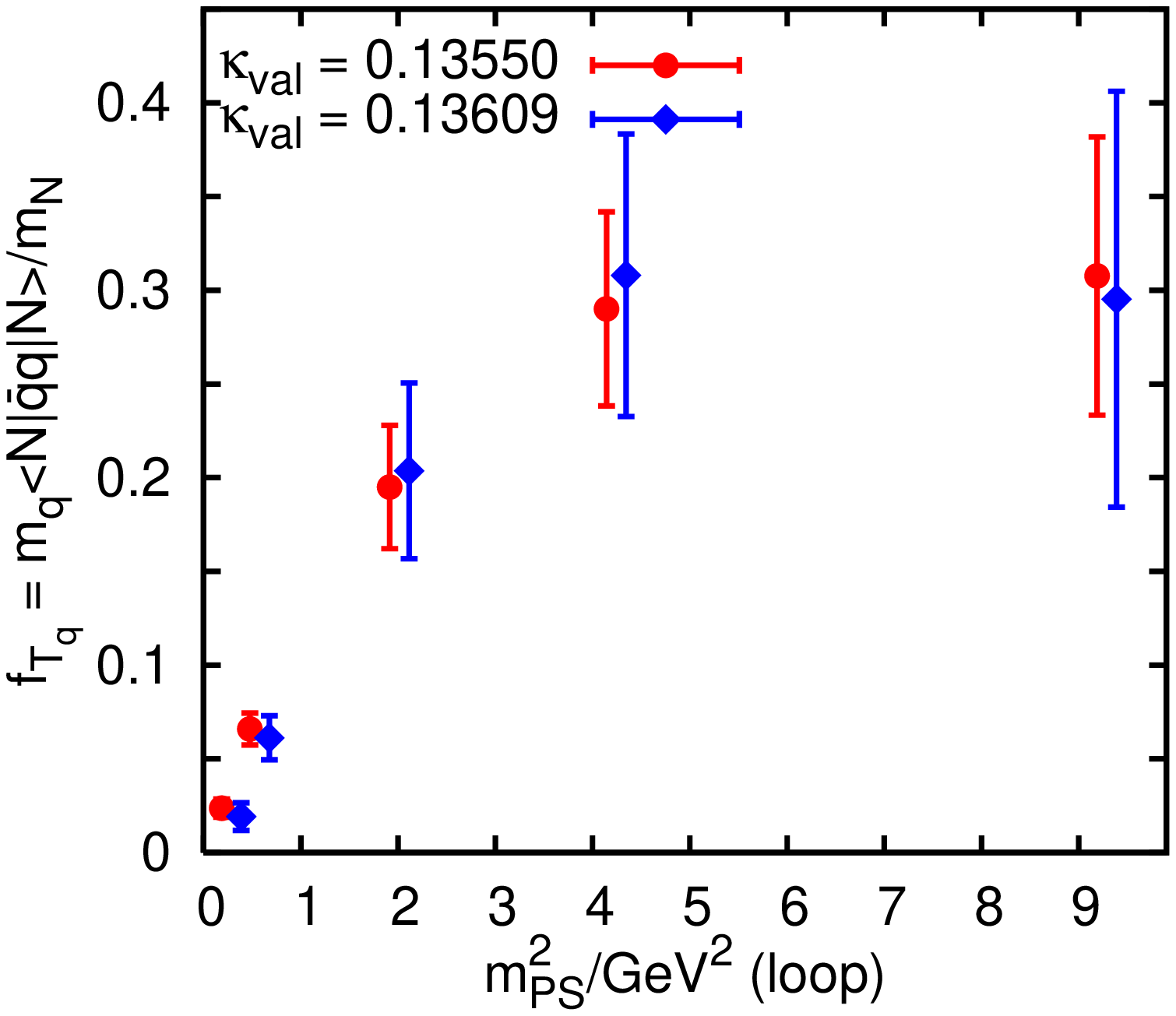,width=.44\textwidth,clip=}{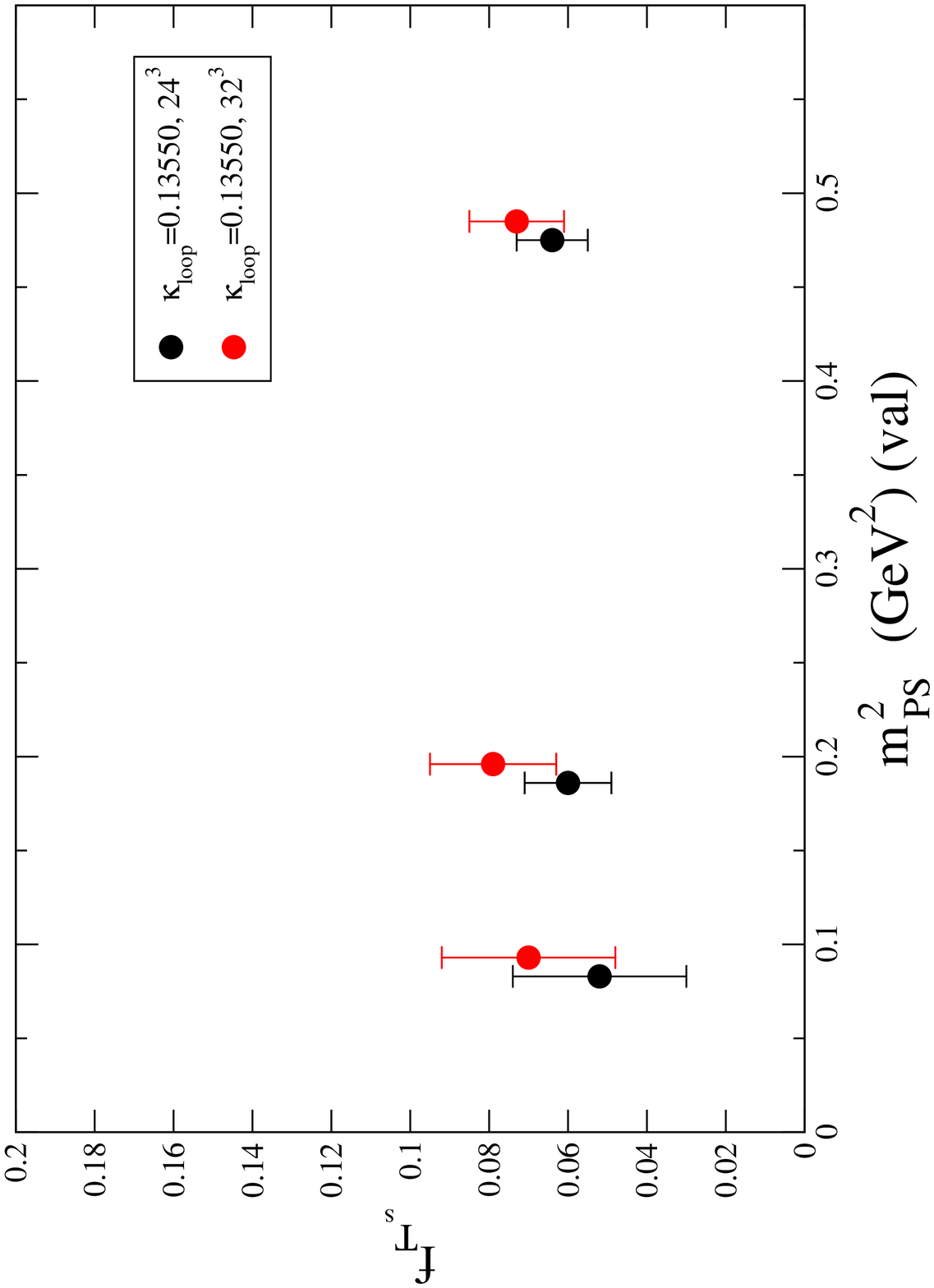,angle=270,width=.51\textwidth,clip=}{The proton mass fractions $f_{T_q}$ for quarks
$q$ of different masses.\label{fig4}}{The strangeness $f_{T_s}$ for different
valence quark masses and volumes.\label{fig5}}

In figure~\ref{fig4} we display our $La\approx 1.8\,$fm results on
$f_{T_q}$ for two different
nucleon valence quark masses $\kappa=\kappa_{\rm strange}$ (red symbols)
and $\kappa=0.13609$ (blue symbols), as functions of the
quark mass $m_q\propto m_{\rm PS}^2$ in the loop. The scheme independent
combination $m_q\langle N|\bar{q}q|N\rangle$ encodes
the contribution of the mass $m_q$ of quark $q$ to the nucleon mass.
This suggests the normalization with respect to the
nucleon mass $m_{\rm N}$.
We note that differences between
the PCAC and the bare quark masses
$m_q=(\kappa^{-1}-\kappa_{c,{\rm val}}^{-1})/(2a)$, where
$\kappa_{c,{\rm val}}$ is obtained from an extrapolation
of $m_{\rm PS}^2$ to zero, only exceed the percent level
for $m_q<m_s$.
The right-most data correspond to
the charm, the data near $m_{\rm PS}^2\approx 0.48\,\,\mbox{GeV}^2$
to the strange quark. $f_{T_q}$ approaches zero
as $m_q\rightarrow 0$ and saturates at $f_{T_q}
\approx 1/3$ for $m_q\gtrsim (2/3)m_c$.
Obviously the charm, bottom or top quarks
cannot each be made responsible for one third of
the proton's mass. Moreover,
in the heavy quark limit, $\langle N|\bar{q}q|N\rangle\sim
\langle N|GG|N\rangle/m_q$, so that mixing with the gluonic
matrix element needs to be considered carefully. 

In figure~\ref{fig5} we focus on the volume and
quark mass dependence of $f_{T_s}$.
The chiral behaviour, varying the valence pseudoscalar mass
from 690\,MeV down to the 290\,MeV that correspond to the sea,
is well fitted by a constant. So, hopefully,
the dependence on the sea quark mass will be weak as well.
$f_{T_s}$ appears to increase on the larger volume,
see also figure~\ref{fig3},
but to exclude this to be just a statistical fluctuation, we will
have to wait for the $L=40$ analysis. The value obtained 
on the larger volume for the lightest nucleon mass reads,
\[
f_{T_s}=\frac{m_s\langle N|\bar{s}s|N\rangle}{m_{\rm N}}=0.070\pm 0.022\,.
\]
Other recent direct calculations resulted in the values
$f_{T_s}=0.34(5)$ for anisotropic
$n_F=2$ Wilson fermions~\cite{Babich:2009rq}
and $f_{T_s}=0.015(28)$ for $n_F=2$ overlap fermions, fixed
to zero topology~\cite{Takeda:2009ga}. An indirect
determination with rooted staggered fermions,
combining chiral condensate and nucleon
two point function data with the
Hellmann-Feynman theorem, suggested
$f_{T_s}=0.063(11)$~\cite{Toussaint:2009pz}.
\FIGURE
{
\rotatebox{270}{\includegraphics[height=.9\textwidth,clip]{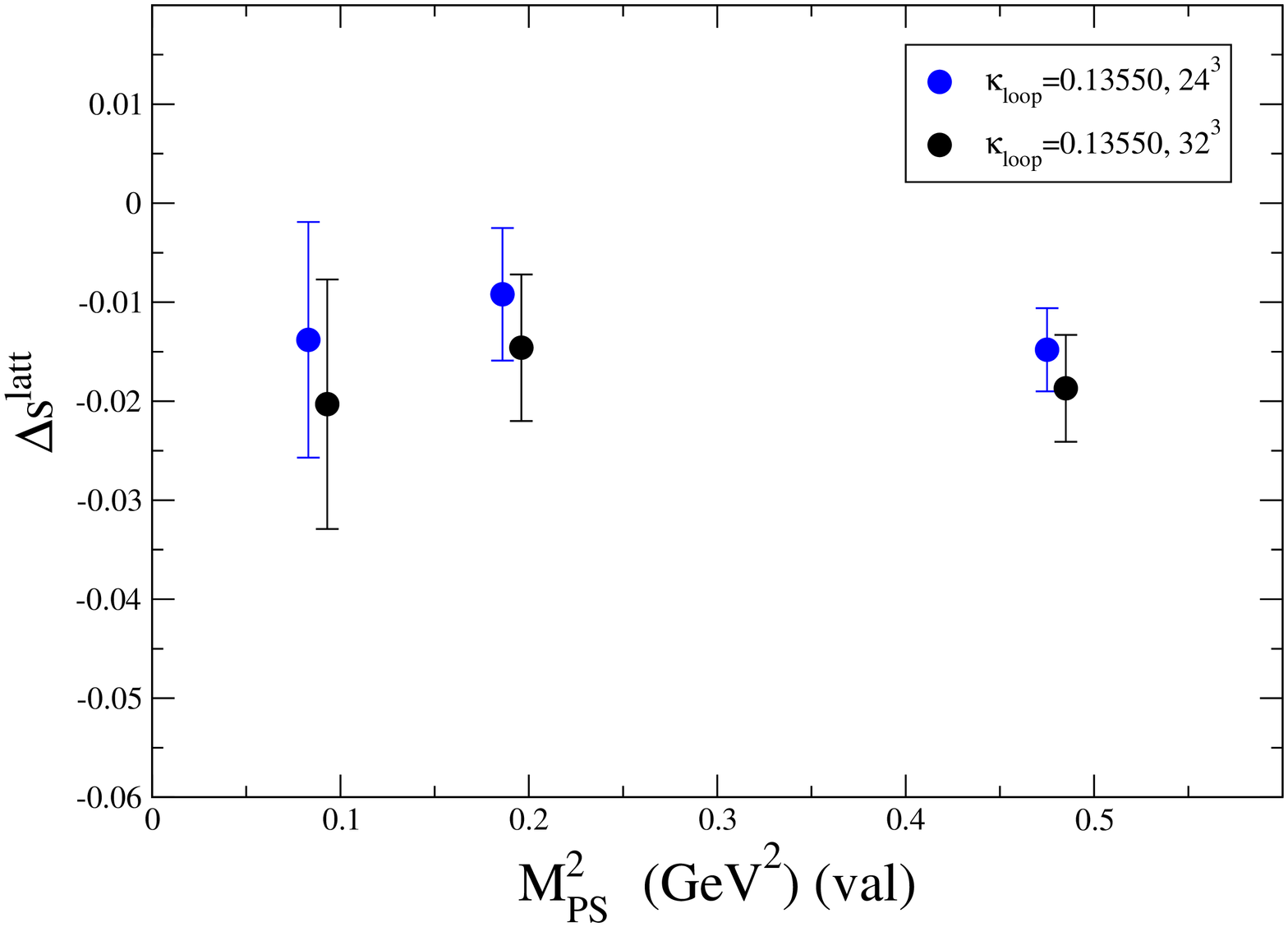}}
\caption{
Valence quark mass dependence of $\Delta s^{\rm latt}$
on the $La=24a\approx 1.8\,$fm and
$La=24a\approx 2.4\,$fm lattices.
\label{fig6}}}

The valence quark mass and volume dependence
of $\Delta s$ is displayed in figure~\ref{fig6}.
Both appear to be mild, with the tendency of a bigger
$-\Delta s$ on the larger volume, in
particular for the lightest nucleon mass. We quote the
value,
\[\Delta s^{\rm latt}=-0.020\pm 0.013\,,
\]
that we obtain on the large volume for the lightest
nucleon mass. Note that this number applies to
the lattice scheme and needs to be multiplied
by a renormalization factor that we expect to be
close to $0.76$ for a conversion into the
$\overline{MS}$ scheme.
Another recent study yielded
$\Delta s = -0.0064(24)$~\cite{Babich:2009rq},
employing $n_F=2$ anisotropic Wilson fermions.
We remark that our value at $\kappa_{\rm val}=\kappa_{\rm strange}$
reads $-0.0187(54)$ and, therefore, differs
from zero by 3.5 standard deviations. 

\section{Outlook}
At present, the $La\approx 3\,$fm $40^3\times 64$ volumes
are being analyzed.
Disconnected contributions to form factors~\cite{Doi:2009vj}
at non-vanishing momentum transfer and to moments of parton distribution
functions are also of big phenomenological interest and will
obviously extend the present study.
The long term goal of this project is to go to large volumes
at the physical sea quark mass.

\acknowledgments
S.~Collins  acknowledges support from the
Claussen-Simon-Foundation (Stifterverband f\"ur die Deutsche
Wissenschaft). This work was supported by the
EC HadronPhysics2 Integrated Infrastructure
Initiative and the DFG
Sonderforschungsbereich/Transregio 55. Computations were
performed on Regensburg's Athene HPC cluster and on the
BlueGene/P (JuGene) and the Nehalem Cluster (JuRoPA) of the J\"ulich
Supercomputer Center.

\end{document}